\documentclass[12pt,preprint]{aastex}
\def\spose#1{\hbox to 0pt{#1\hss}}
\newcommand\lsim{\mathrel{\spose{\lower 3pt\hbox{$\mathchar"218$}}
     \raise 2.0pt\hbox{$\mathchar"13C$}}}       
\newcommand\gsim{\mathrel{\rspose{\lower 3pt\hbox{$\mathchar"218$}}
     \raise 2.0pt\hbox{$\mathchar"13E$}}}

\def\farcm{\hbox{$.\mkern-4mu^\prime$}}
\def\farcs{\hbox{$.\!\!^{\prime\prime}$}}  

\newcommand{\gpm}[3]{$#1^{+#2}_{-#3}$}

\received{2002 June 25}
\begin{document}

\title{Hubble Space Telescope and Ground-Based Optical and
Ultraviolet Observations of GRB\,010222}

\author{T.~J.~Galama\altaffilmark{1}
D.~Reichart\altaffilmark{1}, 
T.~M.~Brown\altaffilmark{2}, 
R.~A.~Kimble\altaffilmark{3}, 
P.~A.~Price\altaffilmark{1},
E.~Berger\altaffilmark{1}, 
D.~A.~Frail\altaffilmark{4}, 
S.~R.~Kulkarni\altaffilmark{1},
S.~A.~Yost\altaffilmark{1}, 
A.~Gal-Yam\altaffilmark{5},
J.~S.~Bloom\altaffilmark{1},
F.~A.~Harrison\altaffilmark{1}, 
R.~Sari\altaffilmark{6},
D.~Fox\altaffilmark{1}, 
S.~G.~Djorgovski\altaffilmark{1}}

\altaffiltext{1}{Division of Physics, Mathematics and Astronomy,
 California Institute of Technology, MS~105-24, Pasadena, CA 91125}

\altaffiltext{2}{Space Telescope Science Institute, 3700 San Martin
Drive, Baltimore, MD 21218}

\altaffiltext{3}{Laboratory for Astronomy and Solar Physics, NASA Goddard Space Flight Center, Code 681, Greenbelt, MD 20771}

\altaffiltext{4}{National Radio Astronomy Observatory, P.O. Box 0,
Socorro, NM 87801}

\altaffiltext{5}{School of Physics and Astronomy, Tel Aviv University,
Tel Aviv 69978, Israel}

\altaffiltext{6}{California Institute of Technology,
 Theoretical Astrophysics  103-33, Pasadena, CA 91125}

\begin{abstract}
We report on {\em Hubble Space Telescope} WFPC2 optical and STIS
near ultraviolet MAMA observations, and ground-based optical
observations of GRB\,010222, spanning 15 hrs to 71 days. The
observations are well-described by a relativistic blast-wave model
with a hard electron-energy distribution, $p = 1.57_{-0.03}^{+0.04}$,
and a jet transition at $t_* = 0.93_{-0.06}^{+0.15}$ days. These
values are slightly larger than previously found as a result of a
correction for the contribution from the host galaxy to the late-time
ground-based observations and the larger temporal baseline provided by
the {\em Hubble Space Telescope} observations. The host galaxy is
found to contain a very compact core (size $<0.25^{''}$) which
coincides with the position of the optical transient. The STIS
near ultraviolet MAMA observations allow for an investigation of the
extinction properties along the line of sight to GRB\,010222. We find
that the far ultraviolet curvature component $c_4$ is rather
large. In combination with the low optical extinction $A_{\rm V} =
0.110_{-0.021}^{+0.010}$ mag, when compared to the Hydrogen column
inferred from X-ray observations, we suggest that this is evidence for
dust destruction.
\end{abstract}


\altaffiltext{1}{Division of Physics, Mathematics and Astronomy, 105-24, 
California Institute of Technology, Pasadena, CA, 91125.} 

\section{Introduction}

\subsection{GRB\,010222}

On February 22, at 7:23:30 UT (2001) the bright GRB\,010222, was
observed by the BeppoSAX Wide Field Camera \#1 (WFC; Piro
2001a\nocite{pir01b}). Within the 2\farcm5 error radius an optical
transient counterpart was identified at $\alpha = $ 14$^{\rm
  h}$52$^{\rm m}$12\farcm55, $\delta = $
+43$^{\circ}$01$^{'}$06\farcs2 (J2000; with an uncertainty of
0\farcs2) \citep{hen01a,hv01,mkg+01}. Soon this was followed by
detections at X-ray \citep{pir01c}, mm \citep{fpm+01} and radio
\citep{bd01} wavelengths.

Absorption lines yielded a redshift of $z = 1.4768 \pm 0.0002$
\citep{gpj+01,jpg+01,bdh+01,cdk+01,mpp+01a} for the highest redshift
absorber. This system shows kinematic substructure, with two distinct
systems at $z_1 = 1.47667 \pm 0.00005$ and $z_2 = 1.47755 \pm
0.00005$, corresponding to a restframe velocity separation of 106 km
s$^{-1}$, typical for internal motions in galaxies \citep{cdk+01}. The
equivalent widths of the lines are unusually strong in comparison to
metallic line absorbers seen in the spectra of quasars, indicating a
high column density of gas, either suggesting a star-forming region
environment or that the GRB occurred in the disk of the galaxy (or
both). The highest redshift absorber is therefore likely the host
galaxy of the GRB itself \citep{jpg+01}.  In addition two foreground
absorbing systems are detected \citep{bdh+01,cdk+01}. These systems
are typical for the metallic line absorbers at comparable redshifts.
Further analysis of the two foreground absorbing systems is reported
in \citet{hal+02} and \citet{skv+02}. Frail et
al.~(2002)\nocite{fbm+02} detected excess submillimeter emission
toward GRB\,010222 which they interpret as originating from a
starburst host galaxy with SFR$\sim$500 M$_\odot$ yr$^{-1}$, much of
which is obscured at optical wavelengths.

Optical observations have been presented by \citet{cpa+01},
\citet{ltv+01}, \citet{mpp+01a}, \citet{ssb+01}, and \citet{sgj+01}.
Although there is good agreement among these authors that the optical
light curves exhibit an achromatic break about 0.5 days after the
burst, no consensus has been reached regarding the origin of this
steepening.  Several authors have interpreted the optical afterglow in
terms of the standard relativistic blast wave model \citep{spn98}.  In
this case the break results from a collimated (jetted) outflow
\citep{sph99,rho99}; the synchrotron cooling frequency is below the
optical passband and the index $p$ of the electron energy distribution
is unusually hard, $p \sim 1.4$ \citep{cpa+01,ssb+01,sgj+01,pk02}.
Other authors prefer to interpret the break as a dynamical transition
to non-relativistic expansion \citep{ika+01,mpp+01a}, requiring a very
energetic burst ($\sim 10^{54}$ erg) expanding into a dense ($\sim
10^{6}$ cm$^{-3}$) circumburst medium. More recently, Bj\"ornsson et
al. (2002)\nocite{bhp+02} have argued in favor of the jet model but
invoke continuous energy injection to explain why the standard model
may erroneously be deriving a hard electron energy index.


Here we report on {\em Hubble Space Telescope} WFPC2 and STIS near-UV
MAMA observations, and ground-based optical observations. We interpret
the resulting lightcurves in the context of a relativistic blast-wave
model with a hard electron-energy distribution. The relatively large
spectral baseline provided by the STIS NUV MAMA observations allows an
investigation of the extinction properties along the line of sight to
GRB\,010222. In \S \ref{sec:data} we present the observations and
the data analysis. In \S \ref{sec:analysis} we present the optical
lightcurves, model fits to the data and investigate the extinction
properties along the line of sight. In \S \ref{sec:conc} we discuss
the results and conclude.

\section{Observations and Data Reduction \label{sec:data}}

\subsection{HST STIS near-UV MAMA Observations \label{sec:stis}}

On 2001 Feb 26.15 UT, the optical transient (OT) was imaged with the
{\em Space Telescope Imaging Spectrograph} (STIS) using the near-UV
MAMA and the F25QTZ filter.  This filter spans 1450--3500~\AA\ (thus
excluding terrestrial airglow lines of Ly$~\alpha$ and \ion{O}{1}),
but little source flux is expected below the redshifted Lyman limit of
the OT (2253~\AA).  We obtained six exposures spanning 2.42 -- 4.64
hours UT, with exposure times of 800--866 sec.  The OT was placed near
the center of the detector in each exposure, but dithering of
5--16~pixels placed the OT at a different location in each frame, to
smooth over small-scale variations in sensitivity.

The raw STIS exposures were reduced using the CALSTIS package of the
STIS Instrument Definition Team \citep{lin99}.  Using IDL, we
performed aperture photometry on each frame, assuming a source
aperture of 10 pixel radius (0.25$^{"}$) and a sky annulus spanning
radii 30--50 pixels.  The OT is well detected in each frame, at a
signal-to-noise ratio $>$ 10.  Summing the net counts in each frame
and dividing by the total exposure time yields a mean count rate of
$0.323\pm 0.012$ cts s$^{-1}$.  There is no evidence for variability
over the two hours of STIS exposures; the count rate measured in each
frame scatters within 1$\sigma$ around the mean rate.  The encircled
energy within the source aperture is 0.856 for near-UV MAMA imaging at
these wavelengths \citep{rob97}; with an aperture correction, the
measured countrate in the STIS bandpass is thus $0.377 \pm 0.014$ cts
s$^{-1}$.  To convert the STIS countrate to flux at a given wavelength
requires the assumption of a spectral energy distribution, a
discussion of which we present in \S \ref{sec:mod}.

\begin{figure}
\plotone{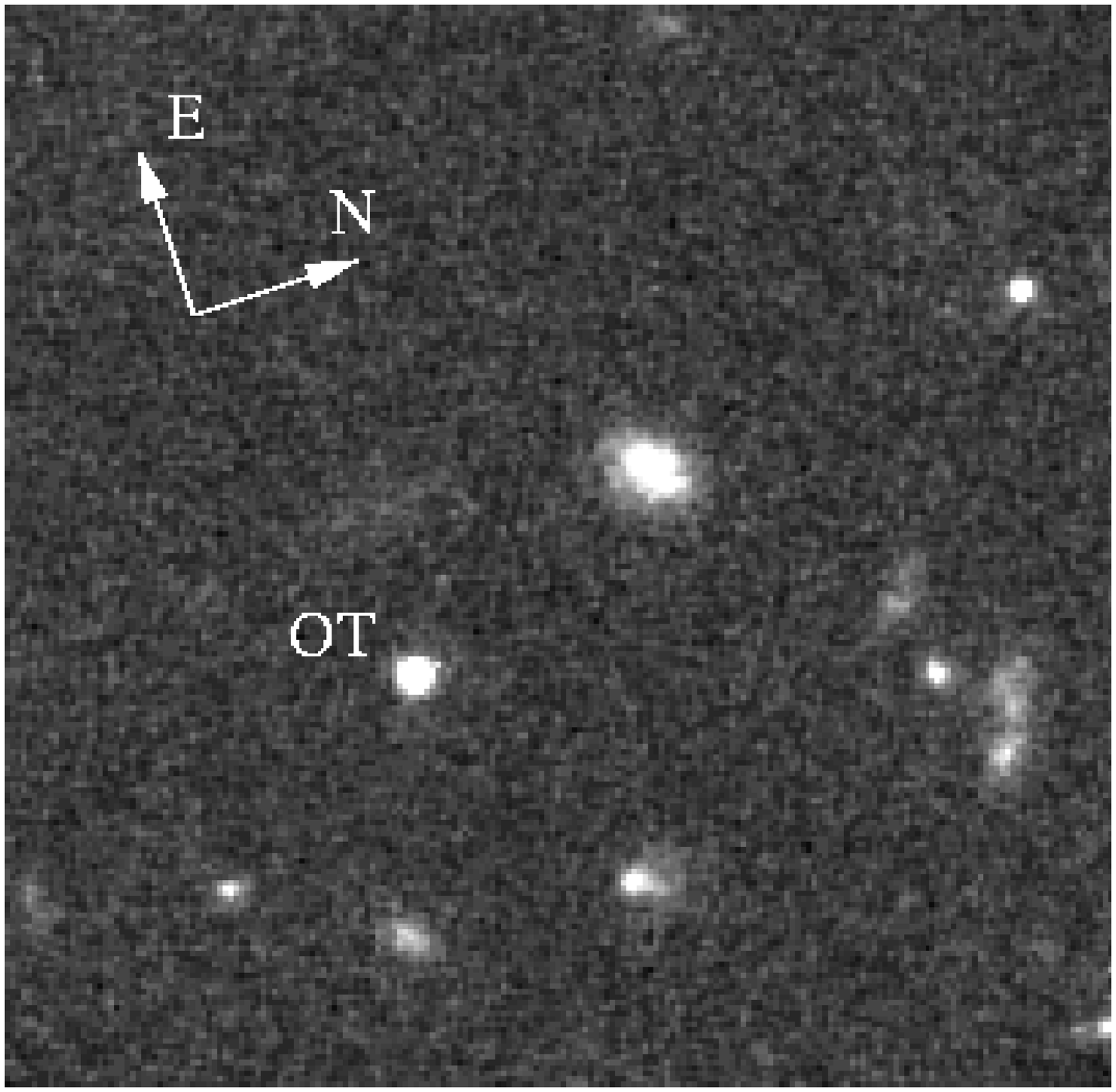}
\caption{F606W drizzled image (each pix is 0.071 arcsec) of a region
around the optical transient (OT). The OT and the Galaxy to the North
East are separated by 3.96 arcsec. \label{fig:f606w}}
\end{figure}

We co-added the individual exposures using the \texttt{drizzle} package
\citep{fh98} to determine if other faint objects are visible in the
STIS image.  The coaddition was done with a mask for hot and bad
pixels, and included a correction for the geometric distortion in the
STIS camera.  Three extended objects lie near the OT; these objects
are also well-detected in the WFPC2 images (see Figure
\ref{fig:f606w}).  We performed aperture photometry on these objects
assuming a source aperture radius of 50 pixels and a sky annulus
spanning 50--70 pixels.  The objects are non-circular in shape, but
this source aperture encloses the detectable flux from each.  The
first source lies 7\farcs7 S of the OT, and has a count-rate of
$0.83\pm 0.06$ cts sec$^{-1}$; the second source lies 7\farcs6 SE of
the OT, with a count-rate of $0.30 \pm 0.06$ cts sec$^{-1}$; the third
source lies 4\farcs2 NNE of the OT, with a count-rate of $0.20 \pm
0.06$ cts sec$^{-1}$.  Without any knowledge about the SEDs, these
count rates cannot be converted to fluxes accurately, but it is worth
noting that in the STIS bandpass, these objects are of a brightness
comparable to that of the OT.

\subsection{HST WFPC2 Observations}

We obtained {\em Wide Field Planetary Camera 2} (WFPC2) HST
observations in the F450W, F606W and F814W passbands at four epochs
as part of an {\em HST} cycle 9 program. These observations took place
between 2001 February 28.66 UT and May 4.73 UT (6.35--71.42 days after
the GRB). The transient was positioned at the WFALL position on
WFPC2 CCD \#3. A log of the observations is provided in Table
\ref{tab-hst}.  

The F450W images were combined using the STSDAS task \texttt{crrej}.
The F606W and F814W images at each epoch were observed at two offsets,
offset by +2.5, +2.5 pixels in x and y. These images were combined and
cosmic-ray rejected using the \texttt{drizzle} technique
\citep{fh98}. The drizzled images have pixels half the area of the
original WFPC2 data. We determined the mag of the optical transient
(OT) in a 2 pixel aperture radius (2.83 pixels for the drizzled
images) and determined the aperture correction from 2 to 5 pixels (7.1
pixels for the drizzled images) to obtain the corresponding 5 pixel
radius magnitude. We corrected the magnitudes for geometric distortion
of the images (\citet{hbc+95}; a correction of 0.024 magnitudes) and
for non-optimal Charge Transfer Efficiency (CTE; less than 0.023
magnitudes). Next, we calibrated the WFPC2 data using the zeropoints
and color transformations to the Johnson Coussins system in Table 10
of \citet{hbc+95}. These magnitudes are given in Table \ref{tab-hst}
and \ref{tab-hst2}. The quoted errors are statistical only; we
estimate the uncertainty in the absolute calibration to be about 0.05
in $B$, $V$, $R$ and $I$.

The F606W drizzled image of a region around the OT of Feb 28.75 UT can
be seen in Fig. \ref{fig:f606w}. The three objects visible in the STIS
NUV-MAMA images are also identified in the WFPC2 images although one
of the objects falls on the edge of the CCD. The galaxy 3.96 arcsec
North East of the OT is the only one of those three that is
displayed. In addition several other galaxies are seen.

We obtain curves of growth to measure the magnitudes of the host in
the WFPC2 images; correcting for the contribution of the OT and the
contribution of the galaxy in a 2 pixel radius (see
\S \ref{sec:mod}), we find $B_{\rm gal}$ = \gpm{25.71}{0.32}{0.24} mag,
$V_{\rm gal}$ = \gpm{26.18}{0.15}{0.13} mag, $R_{\rm gal}$ =
\gpm{26.42}{0.27}{0.21} mag, and $I_{\rm gal}$ =
\gpm{25.66}{0.31}{0.24} mag.

\begin{figure}
\plotone{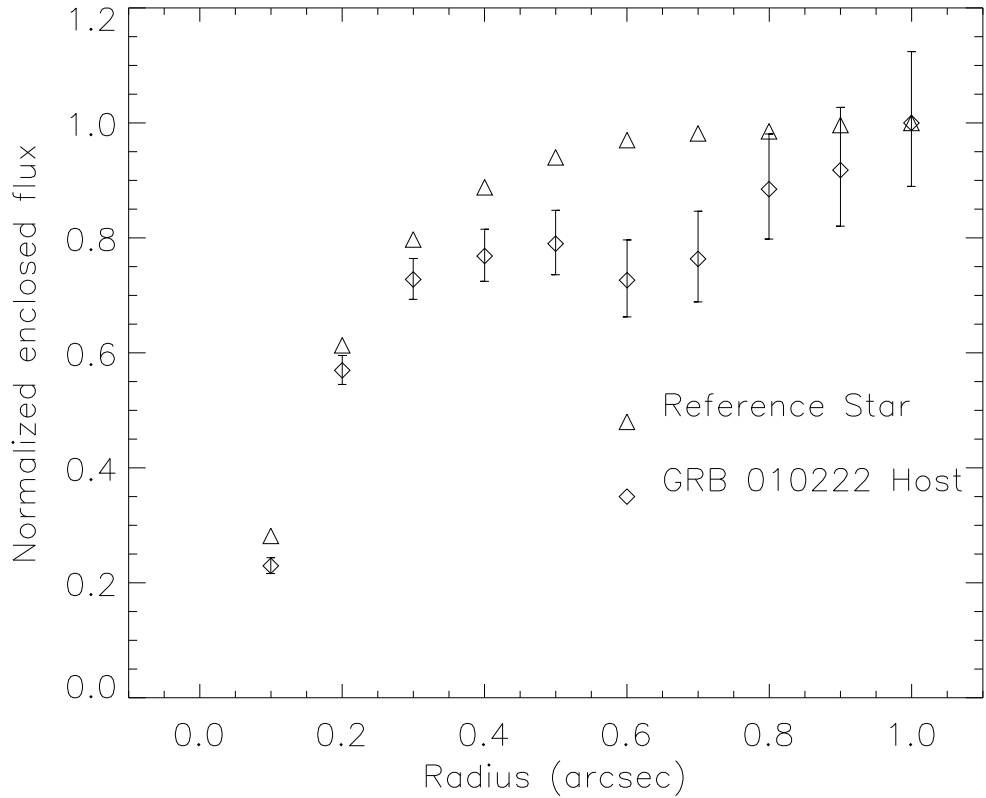}
\caption{Emission from a region around the optical transient (OT) in
the Sept. 8 HST F606W drizzled image as a function of extraction
radius (diamonds) compared to a reference star (triangles). The
vertical scale is the enclosed flux normalized to unity at 1\farcs0
radius. The figure shows evidence for a compact knot of emission
(presumably the core of the host galaxy) and evidence for a low-level
extended emission component, when compared to the instrument point
spread function.
\label{fig:psf}}
\end{figure}

In the final epoch HST images the flux measured at the OT position is
dominated by the host galaxy emission. Figure \ref{fig:psf} shows the
WFPC2 F606W fluxes extracted with increasing apertures, and normalized
to unity at 1\farcs0, for the host emission (diamonds) compared to a
reference star (triangles). The figure shows evidence for a very
compact emission component, 30--50 \% of which can be attributed to
the optical transient, and an additional low-level extended emission
component. 

A cross-correlation of epoch 1 and epoch 6 F606W imaging data allows
for us to place the OT position, measured using IRAF/CENTER/OFILTER in
epoch 1, atop the apparent host galaxy in epoch 6 (see \citealt{bkd01}
for details). The host center was determined using a centroid
algorithm assuming that the light in epoch 6 is dominated by the host.
We determine the offset of the optical transient from the host galaxy
to be 42.9 $\pm$ 5.6 mas East and 7.8 $\pm$ 5.8 mas South, i.e., a
total offset of 43.6 $\pm$ 5.6 mas.  This amounts to an offset of 390
$\pm$ 51 pc in projection at the redshift of GRB 010222 and places the
offset of GRB 010222 in the 20\% percentile of observed GRB offsets
to-date \citep{bkd01}. Note, that since the offset is smaller than the
resolution of final images, if the OT contributes some of the light to
the core of the host in epoch 6, probably not more than 30\%, then the
true offset could be larger by not more than a factor of $\sim$1.3.


\subsection{Ground-Based Optical Observations}

Following the identification of the optical afterglow, we commenced
multi-colour observations with the Wise Observatory 1-m telescope. The
Wise observations were made with the SITe 2k x 4k CCD, using on-chip
2x2 binning, resulting in a pixel scale of 0.8$^{"}$/pixel.  The
observations lasted until about 3 days after the burst. Observations
from the Palomar Observatory were hampered by poor weather, but we
obtained observations 4.2 hours after the event in a Sloan $r'$ filter
with the 200-inch Hale telescope using the {\em Large Format Camera}
(LFC) and in B, V, R and I on 2.468 March 2001 UT using the Jacobs
Camera (JCAM).  The LFC consists of six 2k$\times$4k SITe, thinned,
backside illuminated CCDs, placed in a symmetric cross shaped pattern,
providing a roughly 24 arcmin diameter field of view (FOV). JCAM is a
dual-CCD imager with a 3.2 arcmin diameter FOV \citep{bkd+02}.  The
observations were reduced in the standard manner, and photometered
relative to several field stars calibrated by \citet{hen01b}.  The
errors in the table reflect the statistical and systematic
uncertainties, the latter of which were sizable for the JCAM and LFC
observations because of the lack of suitable secondary standards which
did not allow us to fit reliable colour terms.  We therefore
photometered the afterglow relative to field stars with colours
similar to the OT, without applying any colour correction.  A log of
the observations is presented in Table \ref{tab-opt}.

\section{Afterglow Model \label{sec:analysis}}

\subsection{The Optical Lightcurves \label{sec:opt}}

We suplement the HST WFPC2 and ground-based optical observations from
Tables \ref{tab-hst2} and \ref{tab-opt} with optical and infrared
observations from \citet{cpa+01}, \citet{ltv+01}, \citet{mpp+01a},
\citet{ssb+01}, \citet{sgj+01}, \citet{wkk+01}, and with those
reported in several GRB Coordinate Network (GCN) circulars
\citep{oro01,mdc01,omh+01,vei01a,vei01b,hfg+01}.  We restricted the
dataset to observations for which magnitudes were reported relative to
field stars and converted those to the calibration of \citet{hen01b}
and corrected for a small foreground Galactic extinction
($E_{B-V}=0.023$ mag from Schlegel et al. 1998\nocite{sfd98}).  The
infrared measurements of the OT by \citet{mpp+01a} were taken in poor
seeing (3$^{''}$--4$^{''}$), and the authors therefore used an
aperture of 5$^{''}$ in radius. Keck K-band observations presented in
\citet{fbm+02} show that there are several other (relatively bright)
NIR sources nearby. Within this aperture we measure $K = 17.57 \pm
0.11$; very comparable to what is measured by \citet{mpp+01a}. We
conclude that the NIR measurements of \citet{mpp+01a} are severely
contaminated (thus explaining the flat NIR light curve) and we do not
include these measurements in our dataset. Because the late-time
ground-based optical data is also significantly contaminated by the
host galaxy we have not included two R-band data points 10 days after
GRB\,010222. We converted the Sloan Digital Sky Survey u$^{'}$,
g$^{'}$, i$^{'}$, r$^{'}$, z$^{'}$ observations \citep{ltv+01} to U,
B, V, R, I using the transformations of \citet{fig+96}, assuming no
fading over the course of the 2.5m observations (predicted fading is
0.005 mag, which is negligible), and a constant colour for the 1.5-m
observations of (g$^{'}$-r$^{'}$) = 0.28 $\pm$ 0.03 mag.  An
additional 3\% transformation error was added in quadrature to the
statistical error for these measurements.

\begin{figure}
\plotone{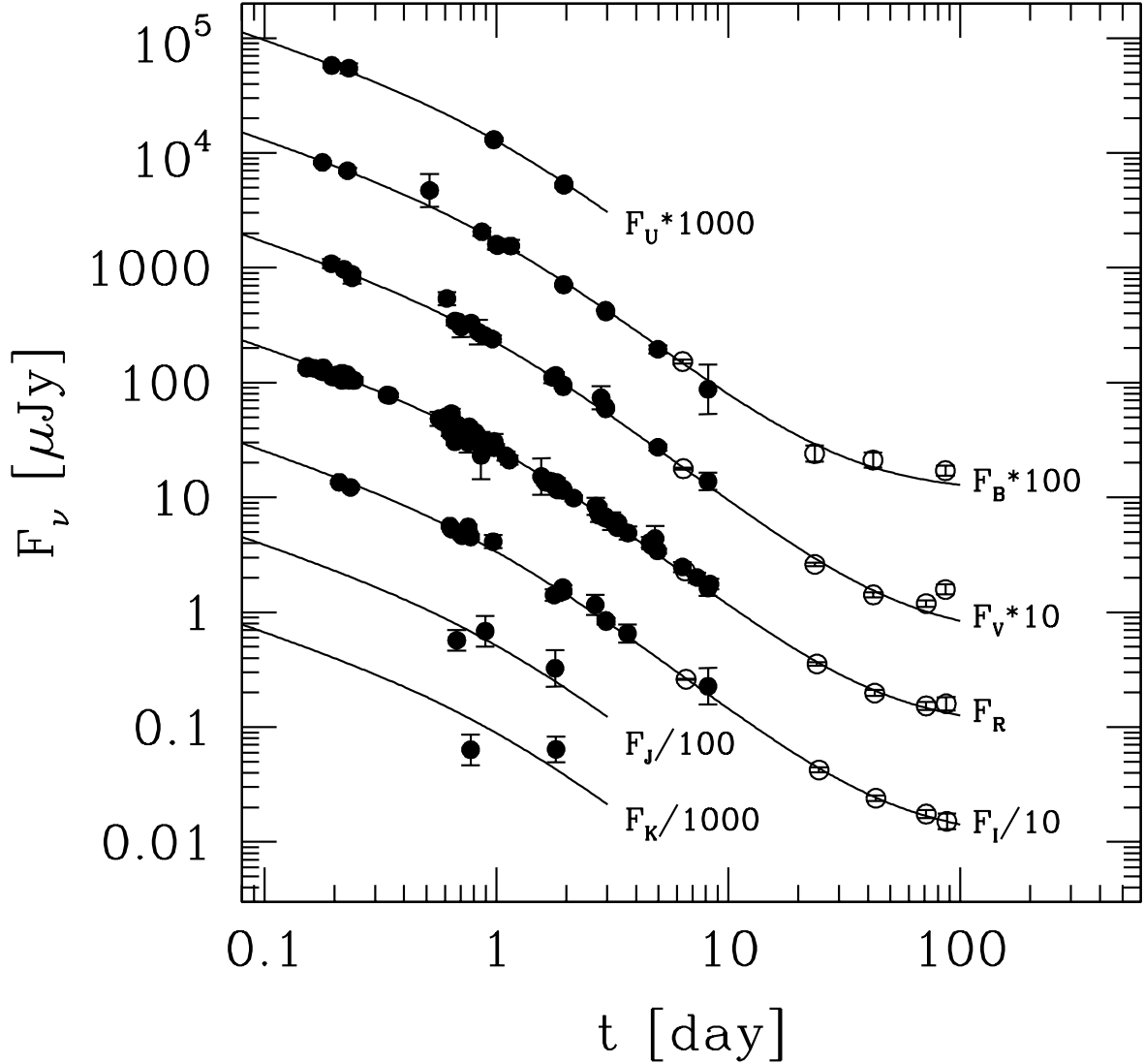}
\caption{The UBVRIJK band lightcurves of GRB\,010222. The HST data is
  shown by open symbols while the ground-based data is shown with
  filled symbols. Shown are fits of Eq. \ref{Fnu} to the data; a
  smoothly broken temporal power law with a break time of $t_* =
  0.93_{-0.06}^{+0.15}$ days and late-time temporal index $\alpha_{2}
  = p = 1.57$, assuming a power law spectrum with index $\beta = p/2$,
  and the full extinction curve of \citet{rei01}. The late-time
  flattening is due to a host contributing to the flux in the HST
  WFPC2 images in a 2 pixel aperture and is modeled by a constant
  contribution. For details see \S \ref{sec:mod}.
\label{fig:opt}}
\end{figure}

In Figure~\ref{fig:opt} we display the optical light curves. We have
added an additional 0.042 mag error in quadrature, the so-called slop
parameter $\sigma$ (see \S \ref{sec:mod}), to reflect uncertainties in
the calibration between different instruments.  The magnitudes have
been converted to flux densities using the transformations of
\citet{bes79} and \citet{bb88}.  The late-time HST (WFPC2)
observations show that the optical light curves, after a shallow break
around 0.7 days, continue roughly at the same decay rate $F_{\nu}
\propto t^{-1.4}$ as previously found by ground-based observations
\citep{cpa+01,mpp+01a,sgj+01,ssb+01}, and show a leveling off to a
constant value in the B, V, R and I band around 25 days (we cannot
establish the presence of a host component in the U and infrared
passbands due to a lack of late-time observations in those bands). The
former has been interpreted as the result of a jet break transition
with a very hard electron energy distribution $p \sim 1.4$
\citep{mpp+01a,ssb+01}. The latter is simply interpreted as due to
emission from the host galaxy of GRB\,010222 in the 2 pixel radius
aperture in which we determined the mag of the OT (WFPC2). This is
consistent with the fact that the source is extended in the WFPC2
images in the later epochs.  Therefore we also have to correct the
ground-based data for the contribution of the host outside the 2-pixel
radius aperture we used for the WFPC2 images. We determine the
difference between a 2 pixel and a large aperture from the HST WFPC2
observations and corrected the ground-based data accordingly.

\subsection{Fits to the Broadband GRB\,010222 Lightcurves \label{sec:mod}}

We now model the optical/NIR photometry and STIS count rate, and
constrain model parameters.  We model the spectrum with a power law
that is extinguished by dust in the host galaxy and our galaxy, and
absorbed by hydrogen in the host galaxy and Ly$\alpha$ forest, and we
model the light curve with a smooth achromatic, broken power law
added to a constant, presumably host galaxy, component:
\begin{equation}
F_{\nu}(t) = e^{-\tau_{\nu}^{MW}} e^{-\tau_{\nu(1+z)}^{Ly\alpha}}\left\{e^{-\tau_{\nu(1+z)}^{host}}F_0\left(\frac{\nu}{\nu_{\rm R}}\right)^{\beta}\left[\left(\frac{t}{t_*}\right)^{-\alpha_1s}+\left(\frac{t}{t_*}\right)^{-\alpha_2s}\right]^{-1/s}+F_{\nu}^{host}\right\},
\label{Fnu}
\end{equation}
where $\tau_{\nu}^{MW}$ is the Galactic extinction curve model of
\citet{ccm89}, $\tau_{\nu(1+z)}^{Ly\alpha}$ is the Ly$\alpha$ forest
absorption model of \citet{rei01}, $\tau_{\nu(1+z)}^{host}$ is the
host galaxy extinction curve and Lyman limit absorption model of
\citet{rei01}, $F_0$ is a normalization parameter, $\nu_{\rm R}$ is
the effective frequency of the R band, $\beta$ is the spectral slope,
$t_*$ is the light curve break time, $s$ is the break smoothness
parameter, $\alpha_1$ is the limiting temporal slope before $t_*$,
$\alpha_2$ is the limiting temporal slope after $t_*$, and
$F_{\nu}^{host}$ is the spectral flux of the host galaxy at frequency
$\nu$.  The STIS bandpass samples 912--1400~\AA\ in the rest-frame,
and so H and H$_2$ absorption lines are expected to be present in the
actual flux distribution; however, a smooth power-law suffices to
describe the flux in the broad STIS bandpass.  We use the full
extinction/absorption model of \citet{rei01}, because \citet{ltv+01}
find a non-standard extinction curve in the source-frame Far
UltraViolet (FUV; see below), and the STIS NUV MAMA transmission
function spans the source-frame FUV, Ly$\alpha$ forest, and Lyman
limit.

Since the source-frame UV extinction curve model has features that are
narrower than most photometric bands, we convolve Equation (\ref{Fnu})
with a boxcar approximation to the appropriate filter function before
fitting it to the optical/NIR photometry.  We have found that failure
to do this can cause one to erroneously identify LMC- and SMC-like
extinction curves when an extinction curve with a large FUV excess
component, possibly related to the fragmentation of grains by the
burst (see below), is actually favored by the data.  In the case of
the STIS measurement, we fit
\begin{equation}
C = A\int_0^{\infty}\frac{F_{\nu}(t_{STIS})}{h\nu}T(\nu)d\nu
\label{C}
\end{equation}
to the measured count rate, where $A$ is the collecting area of the
mirror, and $T(\nu)$ is the transmission function of the mirror and
instrument.

Following previous efforts \citep{cpa+01,ssb+01,sgj+01,pk02} we chose
to fit the data in terms of the standard afterglow model ({\em e.g.,}
Sari, Piran \& Narayan 1998\nocite{spn98}; Sari, Piran \& Halpern
1999\nocite{sph99}). We do not favor an interpretation in which the
break is due to a transition of the blastwave to the non-relativistic
phase \citep{ika+01,mpp+01a}. The same claim has been made for
GRB\,000926 \citep{pgg+01} but the high ambient density causes the
self-absorption frequency to lie far above the radio passband
\citep{hys+01} resulting in an undetectable radio afterglow, contrary
to observations. The jet model with continuous energy injection
\citep{bhp+02} is an interesting alternative but adequate testing of
this model will require a full broadband dataset.

A significant difference between our model fit and others is that
$\alpha_1$ and $\alpha_2$ are not independent but are in fact directly
related to each other by the index $p$ of the electron energy
distribution.  Previous fits to the GRB\,010222 light curves have
treated these two parameters as independent ({\em e.g.,} Stanek et
al.~2001).  We take $\alpha_1 = -(3p-2)/4$, $\alpha_2 = -p$, and
$\beta = -p/2$ (where $p$ is the power law index of the electron
energy distribution), a relation which applies in the standard jet
model when the optical band lies above the synchrotron cooling
frequency \citep{sph99}. This relation may not be obeyed when $1\leq p
\leq 2$ \citep{bha01,pan01,dc01}.  In order to avoid divergence of the
shock energy, an additional cutoff $\gamma_u$ in the electron energy
distribution must be introduced. In general this will modify the
temporal dependence of the decay indices compared to the $p>2$ case.
However, Bhattacharya (2001) has shown that the behavior is identical
when $\gamma_u$ varies in direct proportion to the bulk Lorentz factor
of the shock. Although this is a reasonable assumption, we note that
other solutions are possible depending on what is adopted for the
evolution of $\gamma_u$.

We fit the standard model to the optical/NIR photometry and STIS NUV
MAMA count rate using Bayesian inference ({\em e.g.,} Reichart 2001):
The posterior probability distribution is equal to the product of the
prior probability distribution and the likelihood function.  The
likelihood function is given by
\begin{equation}
{\cal L} = \prod_{i=1}^N\frac{1}{\sqrt{2\pi(\sigma_i^2+\sigma^2)}}\exp{\left\{-\frac{1}{2}\frac{[y(\nu_i,t_i)-y_i]^2}{\sigma_i^2+\sigma^2}\right\}},
\end{equation}
where $N$ is the number of measurements, $y(\nu_i,t_i)$ is the above
described convolution of Equation (\ref{Fnu}) evaluated at the
effective frequency and time of the $i$th measurement if $i < N$, and
$y(\nu_i,t_i) = C$ (Equation \ref{C}) if $i = N$, $y_i$ is the $i$th
measurement in units of log spectral flux if $i < N$, and log count
rate if $i = N$, $\sigma_i$ is the uncertainty in the $i$th
measurement in the same units, and $\sigma$ is a parameter, sometimes
called the ``slop parameter'', that models the small systematic errors
that are unavoidably introduced when data are collected from a wide
variety of sources, and other small sources of error \citep{rei01}.

Many of the parameters of the host galaxy extinction curve model, and
all of the parameters of the Ly$\alpha$ forest absorption model and
Galactic extinction curve model can be constrained {\it a priori}.
The host galaxy extinction curve model of \citet{rei01} is a function
of eight parameters: the source-frame V-band extinction magnitude
$A_{\rm V}$, $R_{\rm V} = A_{\rm V}/E({\rm B}-{\rm V})$, the intercept
$c_1$ and slope $c_2$ of the linear component of the source-frame UV
extinction curve, the strength $c_3$, width $\gamma$, and center $x_0$
of the UV bump component of the extinction curve, and the strength
$c_4$ of the FUV excess component of the extinction curve.  The
Ly$\alpha$ forest model of \citet{rei01} is a function of a single
parameter $D_A$, the flux deficit.  \citet{rei01} finds prior
probability distributions for $R_{\rm V}$, $c_1$, $\gamma$, $x_0$, and
$D_A$, which means that the values of these parameters can be weighted
by fairly narrow distributions, the parameterization of which
sometimes depends on other parameters (in particular $c_2$ and $z$),
{\it a priori}.  We adopt these priors here, which can be thought of
as increasing the degrees of freedom by five.  Also, the Galactic
extinction curve model of Cardelli, Clayton \& Mathis (1989) is a
function of a single parameter $R_{\rm V}^{MW}$.  We adopt a prior for
this parameter that is log-normally distributed with mean $\log{(3.1)}$
and width 0.1, which closely approximates the distribution of values
of this parameter along random lines of sight through the Galaxy
(e.g., Reichart 2001).

\begin{figure}
\plotone{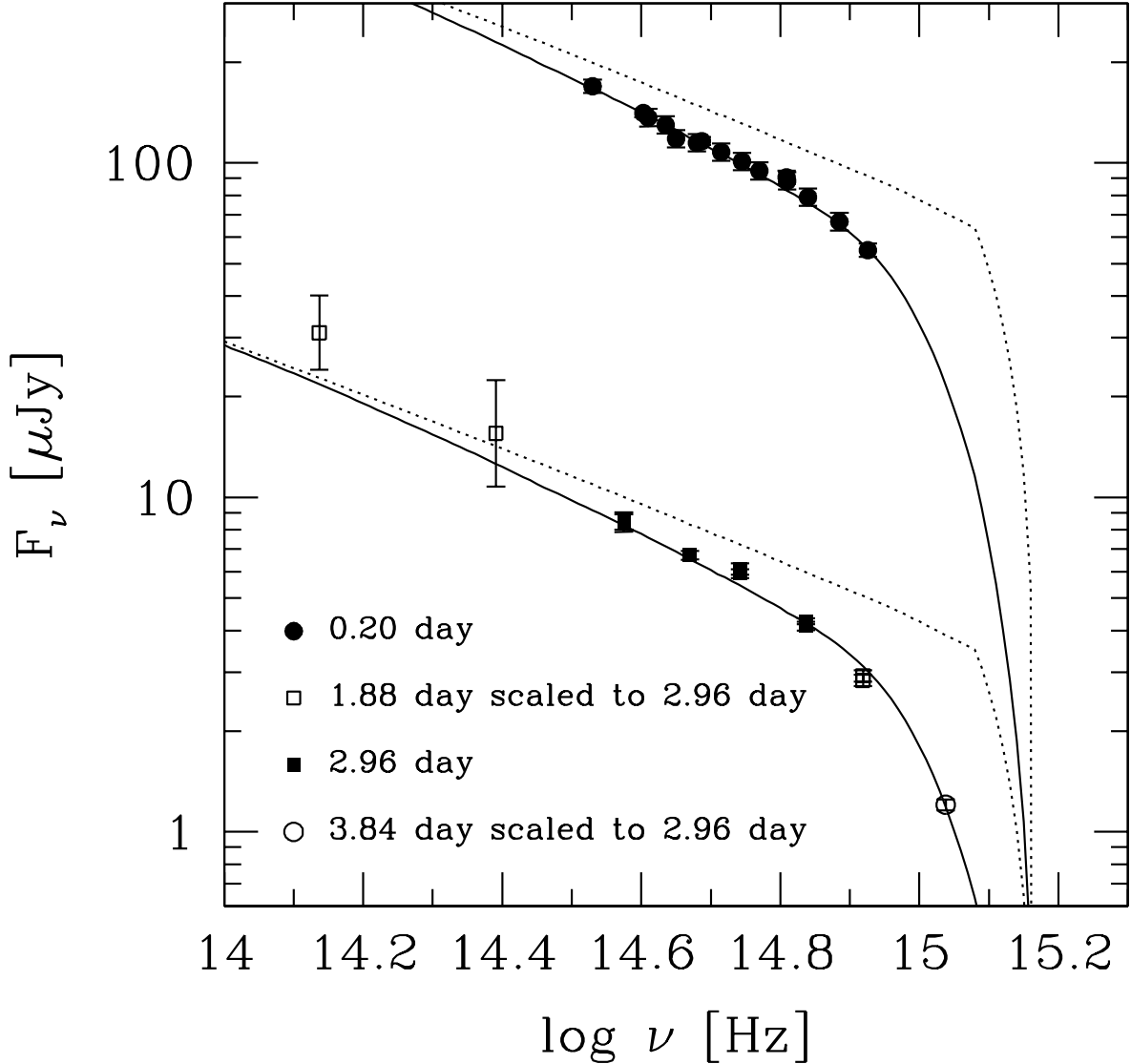}
\caption{Spectral flux distribution at 0.20 days and 2.95 days after
the event. Data are from \citet{ltv+01} and \citet{jpg+01} [filled
circles], \cite{mpp+01a} [filled and open squares], and this work
(STIS NUV MAMA; open circle). Shown is the assumed spectral flux
distribution, a power law with index $\beta$, that is modified by the
Lyman alpha break and forest and small Galactic foreground extinction
(dotted curves).  The index $\beta = p/2$ is determined from a global
fit to all optical data. The fit (solid curves) accounts for
additional extinction in the rest-frame of the OT and employs the full
extinction curve model of \citet{rei01}. See \S \ref{sec:mod} for
details.
\label{fig:spec}}
\end{figure}

The best fit is found by maximizing the posterior.  We find that $F_0
= 60.9_{-5.0}^{+10.2}$ $\mu$Jy, $p = 1.57_{-0.03}^{+0.04}$, $t_* =
0.93_{-0.06}^{+0.15}$ day, $s = 1.51_{-0.08}^{+0.08}$, $F_{\rm
B}^{host} = 0.109_{-0.013}^{+0.014}$ $\mu$Jy, $F_{\rm V}^{host} =
0.103_{-0.011}^{+0.011}$ $\mu$Jy, $F_{\rm R}^{host} =
0.063_{-0.008}^{+0.008}$ $\mu$Jy, $F_{\rm I}^{host} =
0.119_{-0.024}^{+0.027}$ $\mu Jy$, $A_{\rm V} =
0.110_{-0.021}^{+0.010}$ mag, $c_2 = 1.36_{-0.20}^{+0.17}$, $c_3 <
0.12$, $c_4 = 2.02_{-0.39}^{+0.34}$, and $\sigma =
0.042_{-0.005}^{+0.005}$ mag.  We plot the best fit UBVRI light curves
in Figure \ref{fig:opt}, and the best fit spectral flux distribution,
convolved with a canonical filter function, for two epochs in Figure
\ref{fig:spec}.  In Figure \ref{fig:spec}, we plot the effective
spectral flux $1.21\pm0.04$ $\mu$Jy and frequency $1.09\times10^{15}$
Hz of the STIS measurement for our best-fit spectrum.  However, the
uncertainty in our fitted spectrum is not reflected in these values
and uncertainties, and consequently they should not be used in
modeling efforts.  Rather, the measured count rate should be fitted to
using Equation (\ref{C}).

These results are consistent with the results of \citet{ltv+01} for
their assumed spectral slope of $\beta = -0.75$: \citet{ltv+01} find
that $A_V < 0.06$ mag at the 1 $\sigma$ confidence level, and $< 0.27$
mag at the 2 $\sigma$ confidence level, but $> 0$ mag at the 4.3
$\sigma$ confidence level.  We find that $A_V > 0$ mag at the 5.4
$\sigma$ confidence level.  Furthermore, \citet{ltv+01} find that $c_2
= 1.35^{+0.18}_{-0.21}$, and that $c_4 > 0$ at the 2.5 $\sigma$
confidence level, and $c_4 > 1$ -- the largest value previously
observed -- at the 2.1 $\sigma$ confidence level.  We find that $c_4 >
0$ at the 5.3 $\sigma$ confidence level, and that $c_4 > 1$ at the 2.8
$\sigma$ confidence level. A canonical value for $c_4$ is about 0.5,
and $c_4$ ranges between 0 to 1 in the Milky Way, the LMC and the SMC
\citep{rei01}. \citet{ltv+01} suggest that this stronger than expected
FUV excess component of the extinction curve might be due to
sublimation and fragmentation of circumburst dust (e.g, Waxman \&
Draine 2000; Galama \& Wijers 2001): If small (radius $< 300$ ${\rm
  \AA}$) graphite grains, which are probably responsible for the FUV
excess component of the extinction curve (e.g., Draine \& Lee
1984\nocite{dl84}) survive in greater numbers than other grains, this
value of $c_4$ would not be unexpected. This result is seemingly at
odds with the recent work of Perna, Lazzati, \& Fiore (2003) who have
modeled the temporal evolution of the dust properties subject to an
strong X-ray/UV radiation field. They find that the dust destruction
preferentially occurs through sublimation of smaller grains leading to
a flat extinction curve.

\citet{ps95} have measured the relation between optical extinction and
hydrogen column for the Milky Way. Assuming a Galactic relation
between $A_{\rm V}$ and $N_H$, our measurement of $A_{\rm V} =
0.110_{-0.021}^{+0.010}$ mag corresponds to a Hydrogen column of $N_H
= 1.8 \times 10^{20}$ cm$^{-2}$. The hydrogen column density is found
to be $N_H = (1.5 \pm 0.3)\times 10^{21}$ cm$^{-2}$ \citep{ika+01},
corresponding to a restframe column density of $N_H = (1.4 \pm
0.3)\times 10^{22}$ cm$^{-2}$ i.e. 78 $\pm$ 17 times larger than
expected on the basis of the Galactic relation. If the value of $N_H$
derived from {\em Chandra} observations by Bj\"ornsson et al.~(2002)
this ratio is about a factor of two lower.

\section{Discussion and Conclusions \label{sec:conc}}

The late-time optical observations with WFPC2 on HST flatten after
$\sim 25$ days, indicating the presence of a host galaxy; we derive
from the WFPC2 images $B_{\rm gal}$ = \gpm{25.71}{0.32}{0.24} mag,
$V_{\rm gal}$ = \gpm{26.18}{0.15}{0.13} mag, $R_{\rm gal}$ =
\gpm{26.42}{0.27}{0.21} mag, and $I_{\rm gal}$ =
\gpm{25.66}{0.31}{0.24} mag.  We find that the host galaxy emission is
dominated by a very compact core, similar to, for example, the host of
GRB\,970508 \citep{fpg+00}. Similar conclusions about the host galaxy
properties were also reached by Fruchter et al.~(2001)\nocite{fbrl01}
from an independent analysis of these same data.

These late-time optical observations also show that the optical light
curves continued their decay slightly faster; we find a significantly
larger value of the electron-energy index $p = 1.57_{-0.03}^{+0.04}$,
and a later jet-break time $t_{*} = 0.93_{-0.06}^{+0.15}$ day than
previous work \citep{cpa+01,mpp+01a,ssb+01,sgj+01} has reported ($p
\sim 1.4$ and $t_{*} \sim 0.6$ days).  At the same time the quality of
the fit is good (requiring a modest value for the ``slop'' parameter,
$\sigma$), and indicates that a single value of $p$ suffices to
describe the observations. We are not required by the observations to
resort to a more complicated model with, e.g., an index, $p$, that
increases with time (reflected in an increase in decay rate with
time). Our finding of a steeper decay rate may be caused by: (i) our
detection of and fitting for a host galaxy contributing to the
late-time emission, which suggests that previous work requires a small
correction for its contribution, and (ii) the jet transition may not
have been fully developed yet in the early-time observations presented
by previous workers (resulting in their finding a lower $p$). For
accurate determination of the physical parameters of the afterglow
late-time sensitive observations with HST are of great importance.
Our determination of $p$ therefore supersedes that of previous work.

Our STIS NUV measurement indicates that the spectral flux distribution
falls off rapidly toward the ultraviolet; the Far UltraViolet (FUV)
curvature component $c_4$ \citep{rei01} is rather large. This may be
due to dust destruction, where larger grains are preferentially
destroyed.  This provides further support for the idea that the early,
hard radiation from the GRB and its afterglow should destroy dust in
the circumburst environment, carving a path out of the molecular cloud
through which later afterglow light can travel relatively unobstructed
\citep{wd00,fkr01}. 

Additional observational support for dust destruction comes from the
fact that the hydrogen column density is found to be $\sim$ 40-80
times larger than expected from the observed $A_{\rm V}$. This appears
to be a trend in GRB afterglows: the column densities are high and
found to be typical of Galactic giant molecular clouds
\citep{gw01,rp02}, but the optical extinction is found to be small
(factors of 10-100 times smaller). \citet{gw01} have interpreted this
as evidence that GRBs occur in dense star-forming regions and that the
dust is indeed being destroyed.

Also from the possible connection of GRBs with supernovae
\citep{gvv+98,kfw+98,bkd+99,rei99,gtv+00,csg+01,bhj+01,blo+02}, we
would expect that GRBs occur in dense star-forming regions; large
amounts of optical extinction are then naturally expected but only
small amounts are observed \citep{gw01}.

On the other hand, early-time Keck spectroscopic observations of
GRB\,010222 \citep{hal+02} show a two-component system similar to that
of GRB\,000926 \citep{cgh+02}. In GRB\,000926 the two absorption
systems have a velocity separation of 168~km~s$^{-1}$, which is
interpreted as being due to two individual clouds in the host galaxy.
Dust destruction by the GRB and the afterglow is effective only to
$\sim 10-100$ pc \citep{wd00,fkr01} from the GRB, and it is therefore
unlikely that the dust in both clouds would be destroyed. Further, the
two clouds appear to have similar relative metal abundance and dust to
gas ratio. Castro et al. (2002) argue that one of the clouds is
probably associated with the GRB site, and consequently that the burst
was ineffective at sublimating dust, in contradiction to theoretical
expectations (e.g., Waxman \& Draine 2000). In other words, the
explanation for the observed low optical extinctions could also be
that GRB host galaxies typically have low dust to gas ratios ({\em
  cf.,} Pettini et al.~1997\nocite{pet97}). This is a simple and
attractive explanation, based only on the assumption that the dust to
gas ratio was the same for both clouds before the burst occurred.
However, in higher redshift galaxies, such as this one, the dust to
gas ratio might only be high in those clouds that are actively
producing stars, which is presumably the case at the GRB site (i.e.,
dust has not had sufficient time to mix with gas elsewhere in the
galaxy). The burst would sublimate this dust, returning the metals to
the gas phase, and consequently, similar relative metalicities and
dust to gas ratios might also be interpreted as evidence in favor of
sublimation.

\acknowledgements E.~O.~Ofek and J.~Dann are thanked for their help
with observations at the Wise Observatory. We wish to thank Steve
Beckwith and the HST operations staff for facilitating the WFPC2 and
STIS observations. TJG acknowledges support from the Sherman Fairchild
Foundation. FAH acknowledges support from a Presidential Early Career
award. SRK and SGD thank NSF for support of their ground-based GRB
programs. AGJ is a Colton Fellow.  JSB is a Fannie and John Hertz
Foundation Fellow.

\clearpage


\clearpage

\clearpage

\clearpage

\clearpage

\begin{deluxetable}{lclc}
\footnotesize 
\tablecolumns{4} 
\tablewidth{0pc} 
\tablecaption{WFPC2 HST observations of GRB\,010222. Reported magnitudes are for a 5-pixel aperture. \label{tab-hst}}
\tablehead{\colhead{Date (2001, UT)} & \colhead{Filter} &
\colhead{Exposure time (sec)} & \colhead{Magnitude}}
\startdata 
Feb 28.66       & F450W & 2x1100 (1 orbit)  & 23.463 $\pm$ 0.034 \\
Feb 28.75       & F606W & 4x1100 (2 orbits) & 23.029 $\pm$ 0.010 \\
Feb 28.89       & F814W & 4x1100 (2 orbits) & 22.440 $\pm$ 0.012 \\
Mar 17.86       & F450W & 2x1100 (1 orbit)  & 25.48  $\pm$ 0.15 \\
Mar 17.96       & F606W & 4x1100 (2 orbits) & 25.075 $\pm$ 0.029 \\
Mar 18.90       & F814W & 4x1100 (2 orbits) & 24.41  $\pm$ 0.05 \\
Apr 5.40        & F450W & 2x1100 (1 orbit)  & 25.71  $\pm$ 0.16 \\
Apr 5.49        & F606W & 4x1100 (2 orbits) & 25.72  $\pm$ 0.04 \\ 
Apr 6.56        & F814W & 4x1100 (2 orbits) & 25.02  $\pm$ 0.06 \\
May 4.35        & F606W & 6x1000 (3 orbits) & 25.96  $\pm$ 0.06 \\ 
May 4.73        & F814W & 4x1100 (2 orbits) & 25.37  $\pm$ 0.08 \\
Sep 8.38        & F450W & 6x1000 (3 orbits) & 25.94 $\pm$ 0.10 \\
Sep 8.59        & F606W & 6x1000 (3 orbits) & 25.82 $\pm$ 0.09 \\
Sep 9.39        & F814W & 6x1000 (3 orbits) & 25.53 $\pm$ 0.15  \\
\enddata
\end{deluxetable}

\clearpage

\begin{deluxetable}{lcccc}
\footnotesize 
\tablecolumns{5} 
\tablewidth{0pc} 
\tablecaption{Transformed WFPC2 HST magnitudes to Johnson Coussins system. \label{tab-hst2}} 
\tablehead{\colhead{Date (2001, UT)} & \colhead{B} & \colhead{V} & \colhead{R} & \colhead{I}}
\startdata 
Feb 28.75       & 23.544 $\pm$ 0.038 & 23.264 $\pm$ 0.017 & 22.818
$\pm$ 0.015 & 22.418 $\pm$ 0.015 \\
Mar 17.96 & 25.55 $\pm$ 0.16 & 25.35 $\pm$ 0.04 & 24.84 $\pm$ 0.04 &
24.39 $\pm$ 0.05 \\    
Apr 5.49 & 25.69 $\pm$ 0.16 & 26.01 $\pm$ 0.05 & 25.47 $\pm$ 0.06 &
25.01 $\pm$ 0.06 \\
May 4.54 & & 26.20 $\pm$ 0.07 & 25.75 $\pm$ 0.08 &
25.35 $\pm$ 0.08 \\
Sep 8.79 & 25.96 $\pm$ 0.10 & 25.92 $\pm$ 0.10 & 25.72 $\pm$ 0.13 & 25.52 $\pm$ 0.15 \\
\enddata
\end{deluxetable}

\clearpage

\begin{deluxetable}{ccccc}
\footnotesize \tablecolumns{4} \tablewidth{0pc}
\tablecaption{Ground-Based Optical Observations of GRB\,010222. P200
LFC images were obtained with Steidel R filter
\label{tab-opt}}
\tablehead{\colhead{Date (UT)} & \colhead{Passband} & \colhead{Magnitude} & \colhead{Telescope} }
\startdata
Feb 22.948 &    R &     19.387 $\pm$ 0.098 & Wise 1-m \\
Feb 22.959 &    R &     19.567 $\pm$ 0.082 & Wise 1-m \\
Feb 22.970 &    V &     20.052 $\pm$ 0.085 & Wise 1-m \\
Feb 23.063 &    I &     19.102 $\pm$ 0.047 & Wise 1-m \\
Feb 23.072 &    R &     19.676 $\pm$ 0.053 & Wise 1-m \\
Feb 23.086 &    V &     20.093 $\pm$ 0.061 & Wise 1-m \\
Feb 24.020 &    R &     20.859 $\pm$ 0.075 & Wise 1-m \\
Feb 24.052 &    V &     21.279 $\pm$ 0.109 & Wise 1-m \\
Feb 24.084 &    I &     20.582 $\pm$ 0.118 & Wise 1-m \\
Feb 25.085 &    R &     21.607 $\pm$ 0.129 & Wise 1-m \\
Feb 25.136 &    V &     21.721 $\pm$ 0.231 & Wise 1-m \\
\hline
Feb 22.482 &    R &     18.455 $\pm$ 0.033 & P200 LFC \\
Feb 22.487 &    R &     18.481 $\pm$ 0.036 & P200 LFC \\
\hline
Mar 2.468 &     I &     22.57  $\pm$  0.35& P200 JCAM \\
Mar 2.468 &     R &     23.19  $\pm$  0.15& P200 JCAM \\
Mar 2.468 &     V &     23.54  $\pm$  0.17& P200 JCAM \\
Mar 2.468 &     B &     24.15  $\pm$  0.46& P200 JCAM \\
\enddata
\end{deluxetable}

\end{document}